\documentstyle[preprint,aps]{revtex}
\tightenlines
\begin{document}
\draft
\title
{A note on Farey sequences and Hausdorff dimension}
\author
{Wellington da Cruz\footnote{E-mail: wdacruz@exatas.campus.uel.br}} 
\address
{Departamento de F\'{\i}sica,\\
 Universidade Estadual de Londrina, Caixa Postal 6001,\\
Cep 86051-990 Londrina, PR, Brazil\\}
\date{\today}
\maketitle
\begin{abstract}
We prove that the Farey sequences can be 
express into equivalence 
classes labeled by  a fractal parameter 
which looks like a Hausdorff dimension $h$ 
defined within the interval 
$1$$\;$$ < $$\;$$h$$\;$$ <$$\;$$ 2$. The classes $h$ satisfy 
the same properties of the Farey series and for 
each value of $h$ there exists an algebraic equation. 

\end{abstract}

\pacs{PACS numbers: 02.10.Lh, 02.30.Lt\\
Keywords: Farey sequences; Hausdorff dimension; 
Number theory }
%\narrowtext

%\

From considerations about Fractional 
Quantum Hall Effect (FQHE)\cite{R1} we have found a connection 
between a fractal parameter or Hausdorff dimension 
$h$ and the Farey series\cite{R2}. 
Thus, we have the following {\bf theorem}:

{\it The elements of the Farey series belong to 
distinct equivalence classes labeled by a fractal parameter $h$
defined into the interval 
$1$$\;$$ < $$\;$$h$$\;$$ <$$\;$$ 2$, such that these classes satisfy 
the same properties observed for that fractions. Also, 
for each value of $h$ there exists an algebraic equation }.

The fractal parameter $h$ is related to $\nu$ ( an 
irreducible number $\frac{p}{q}$, 
with $p$ and $q$ integers ) as follows

\begin{eqnarray}
\label{e.2}
&&h-1=1-\nu,\;\;\;\; 0 < \nu < 1;\;\;\;\;\;\;\;\;
 h-1=\nu-1,
\;\;\;\;\;\;\; 1 <\nu < 2;\;\nonumber\\
&&h-1=3-\nu,\;\;\;\; 2 < \nu < 3;\;\;\;\;\;\;\;\;
h-1=\nu-3,\;\;\;\;\;\;\; 3 < \nu < 4;\;\nonumber\\
&&h-1=5-\nu,\;\;\;\; 4 < \nu < 5;\;\;\;\;\;\;\;
h-1=\nu-5,\;\;\;\;\;\;\;\; 5 < \nu < 6;\;\\
&&h-1=7-\nu,\;\;\;\; 6 < \nu < 7;\;\;\;\;\;\;\;
h-1=\nu-7,\;\;\;\;\;\;\;\; 7 < \nu < 8;\;\nonumber\\
&&h-1=9-\nu,\;\;\;8 < \nu < 9;\;\;\;\;\;\;\;
h-1=\nu-9,\;\;\;\;\;\;\; 9 < \nu < 10;\nonumber\\
&&etc,\nonumber
\end{eqnarray}

We can extract, for example, the classes

\begin{eqnarray}
&&\left\{\frac{1}{3},\frac{5}{3},\frac{7}{3},
 \frac{11}{3},\cdots\right\}_
{h=\frac{5}{3}},\;\;\;\;\;\;\;\;\;\;\;\;\;\;\left\{\frac{5}{14},
\frac{23}{14},\frac{33}{14},\frac{51}{14},\cdots\right\}_
{h=\frac{23}{14}};\nonumber\\
&&\left\{\frac{4}{11},\frac{18}{11},\frac{26}{11},
 \frac{40}{11},\cdots\right\}_
{h=\frac{18}{11}},\;\;\;\;\;\;\;\;\left\{\frac{7}{19},
\frac{31}{19},\frac{45}{19},\frac{69}{19},\cdots\right\}_
{h=\frac{31}{19}};\nonumber\\
&&\left\{\frac{10}{27},\frac{44}{27},\frac{64}{27},
 \frac{98}{27},\cdots\right\}_
{h=\frac{44}{27}},\;\;\;\;\;\;\;\;\left\{\frac{3}{8},
\frac{13}{8},\frac{19}{8},\frac{29}{8},\cdots\right\}_
{h=\frac{13}{8}};\\
&&\left\{\frac{3}{7},\frac{11}{7},\frac{17}{7},
\frac{25}{7},\cdots\right\}_
{h=\frac{11}{7}},\;\;\;\;\;\;\;\;\;\;\left\{\frac{4}{9},
\frac{14}{9},\frac{22}{9},\frac{32}{9},
\cdots\right\}_{h=\frac{14}{9}}\nonumber\\
&&\left\{\frac{5}{11},\frac{17}{11},\frac{27}{11},
\frac{39}{11},\cdots\right\}_
{h=\frac{17}{11}},\;\;\;\;\;\;\;\;\left\{\frac{6}{13},
\frac{20}{13},\frac{32}{13},
\frac{46}{13},\cdots\right\}_{h=\frac{20}{13}},\nonumber\\
&&\left\{\frac{2}{5},\frac{8}{5},\frac{12}{5},
\frac{18}{5},\cdots\right\}_
{h=\frac{8}{5}},\nonumber
\end{eqnarray}

\noindent and for that we can consider the 
serie $(h,\nu)$

\begin{eqnarray}
&&\left(\frac{5}{3},\frac{1}{3}\right)\rightarrow 
\left(\frac{18}{11},\frac{4}{11}
\right)\rightarrow
\left(\frac{13}{8},\frac{3}{8}\right)\rightarrow
 \left(\frac{8}{5},\frac{2}{5}\right)\rightarrow\\
&&\left(\frac{11}{7},\frac{3}{7}\right)\rightarrow 
\left(\frac{14}{9},\frac{4}{9}\right)
\rightarrow \left(\frac{17}{11},\frac{5}{11}\right)
\rightarrow 
\left(\frac{20}{13},\frac{6}{13}\right) \rightarrow 
\cdots\nonumber
\end{eqnarray} 

\noindent The classes $h$ satisfy all properties of the 
Farey series: 

P1. If $h_{1}=\frac{p_{1}}{q_{1}}$ and 
$h_{2}=\frac{p_{2}}{q_{2}}$ are two consecutive fractions 
$\frac{p_{1}}{q_{1}}$$ >$$ \frac{p_{2}}{q_{2}}$, then 
$|p_{2}q_{1}-q_{2}p_{1}|=1$.

P2. If $\frac{p_{1}}{q_{1}}$, $\frac{p_{2}}{q_{2}}$,
$\frac{p_{3}}{q_{3}}$ are three consecutive fractions 
$\frac{p_{1}}{q_{1}}$$ >$$ \frac{p_{2}}{q_{2}} 
$$>$$ \frac{p_{3}}{q_{3}}$, then 
$\frac{p_{2}}{q_{2}}=\frac{p_{1}+p_{3}}{q_{1}+q_{3}}$.

P3. If $\frac{p_{1}}{q_{1}}$ and $\frac{p_{2}}{q_{2}}$ are 
consecutive fractions in the same sequence, then among 
all fractions\\
 between the two, 
$\frac{p_{1}+p_{2}}{q_{1}+q_{2}}$
 is the unique reduced
fraction with the smallest denominator.
 
For more details about Farey series see\cite{R2}. 
All these properties can be verified for the 
classes considered above as an example. Another 
example is

\begin{eqnarray}
(h,\nu)&=&\left(\frac{11}{6},\frac{1}{6}\right)\rightarrow 
\left(\frac{9}{5},\frac{1}{5}
\right)\rightarrow
\left(\frac{7}{4},\frac{1}{4}\right)\rightarrow
 \left(\frac{5}{3},\frac{1}{3}\right)\rightarrow\\
&&\left(\frac{8}{5},\frac{2}{5}\right)\rightarrow 
\left(\frac{3}{2},\frac{1}{2}\right)
\rightarrow \left(\frac{7}{5},\frac{3}{5}\right)
\rightarrow 
\left(\frac{4}{3},\frac{2}{3}\right) \rightarrow\nonumber\\ 
&&\left(\frac{5}{4},\frac{3}{4}\right) \rightarrow 
\left(\frac{6}{5},\frac{4}{5}\right) \rightarrow 
\left(\frac{7}{6},\frac{5}{6}\right) \rightarrow 
\cdots,\nonumber
\end{eqnarray}

\noindent where the $\nu$ sequence is 
the Farey series of order $6$. Thus, we observe 
that because of the {\it fractal spectrum} (Eq.\ref{e.2}), 
we can write down any Farey series of rational numbers.

In summary, we have extracted from 
considerations about FQHE\cite{R1} a beautiful connection 
between number theory and physics. We have shown that 
the Farey series can be arranged into equivalence 
classes labeled by 
a fractal parameter $h$ which looks like a Hausdorff 
dimension. For each value of $h$ we have an algebraic 
equation derived of the functional equation

\begin{eqnarray}
\xi=\left\{{\cal{Y}}(\xi)-1\right\}^{h-1}
\left\{{\cal{Y}}(\xi)-2\right\}^{2-h},
\end{eqnarray}

\noindent where $\xi$ is an exponential function. Then, 
there exists 
a relation between algebraic equation and Farey series. The 
connection between a geometric parameter related 
to paths 
of particles ( charge-flux system called {\it fractons}\cite{R1,R3} 
 identified with holes of a 
 multiply connected space) defined in the context of 
a two dimensional physical system and rational numbers 
deserve a more deep research.

\end{document}